\documentclass[aps,prl,twocolumn,showpacs,superscriptaddress,amsmath,amssymb, showkeys]{revtex4}

\usepackage{graphicx}
\usepackage{dcolumn}
\usepackage{bm}

\newtheorem{theorem}{Theorem}

\def\smallssp{\def\baselinestretch{1.0}\large\small}

\newlength{\pgmtab}  
\newenvironment{program}{\renewcommand{\baselinestretch}{1}%
\begin{tabbing}\hspace{0em}\=\hspace{0em}\=%
\hspace{\pgmtab}\=\hspace{\pgmtab}\=\hspace{\pgmtab}\=\hspace{\pgmtab}\=%
\hspace{\pgmtab}\=\hspace{\pgmtab}\=\hspace{\pgmtab}\=\hspace{\pgmtab}\=%
\+\+\kill}{\end{tabbing}\renewcommand{\baselinestretch}{\intl}}
\newcommand {\BEGIN}{{\bf begin\ }}

\newcommand {\IF}{{\bf if\ }}

\newcommand {\END}{{\bf end}}
\newcommand {\RETURN}{\mbox{\bf return\ }}

\newcommand {\INPUT}{{\bf input}}
\newcommand {\OUTPUT}{{\bf output}}

\begin{document}

\title{Quantum Mechanical Search and Harmonic Perturbation}

\author{Jie-Hong R. Jiang}
\affiliation{Department of Electrical Engineering, National Taiwan
University, Taipei 10617, Taiwan} \email{jhjiang@cc.ee.ntu.edu.tw}
\author{Dah-Wei Chiou}
\affiliation{Department of Physics, University of California,
Berkeley, CA 94720, USA}
\author{Cheng-En Wu}
\affiliation{Department of Physics, National Tsing Hua University,
Hsinchu 30013, Taiwan}


\begin{abstract}
Perturbation theory in quantum mechanics studies how quantum
systems interact with their environmental perturbations. Harmonic
perturbation is a rare special case of time-dependent
perturbations in which exact analysis exists. Some important
technology advances, such as masers, lasers, nuclear magnetic
resonance, etc., originated from it. Here we add quantum
computation to this list with a theoretical demonstration. Based
on harmonic perturbation, a quantum mechanical algorithm is
devised to search the ground state of a given Hamiltonian. The
intrinsic complexity of the algorithm is continuous and parametric
in both time $T$ and energy $E$. More precisely, the probability
of locating a search target of a Hamiltonian in $N$-dimensional
vector space is shown to be $1/(1+ c N E^{-2} T^{-2})$ for some
constant $c$. This result is optimal. As harmonic perturbation
provides a different computation mechanism, the algorithm may
suggest new directions in realizing quantum computers.
\end{abstract}



\maketitle

\section{\label{sec:intro}Introduction}
Quantum physics can in principle speed up solving the
unsorted-database search problem with a quadratic improvement over
classical algorithms, as was first demonstrated by Grover
\cite{Gro96}. This problem was originally formulated as to
identify a target item in the fewest queries to a black-box
database. An important reformulation by Farhi et al.
\cite{FG98,FGGS00} phrased the problem as to search the target
state with some special eigenvalue of a given Hamiltonian
$\mathcal{H}$, which represents the database. In particular,
constant and adiabatic perturbations were proposed in \cite{FG98}
and \cite{FGGS00}, respectively, for quantum search. These methods
yield the same quadratic speed-up as Grover's construct
\cite{DMV01,RC02}. Moreover, the adiabatic computation is
equivalent to standard quantum computation (in terms of unitary
transformations) \cite{ADK04}. Although some physical
implementations have been demonstrated in realizing quantum search
algorithms, e.g. \cite{VSS00,PZFFLG02,SvD03}, they may not be
scalable to solve large problem instances without some fundamental
breakthroughs \cite{Pre98}. Searching alternative computation
models may suggest new ways of building quantum computers.

Despite the success of the constant and adiabatic perturbations in
quantum search, the applicability of perturbations based on fast
time-varying Hamiltonians remains an open problem. This paper
exploits harmonic perturbation for quantum computation. A new
computation model is proposed, inspired by the well-studied
harmonic perturbation of two-state systems in quantum mechanics.
By preparing a system in one or the other of its two states
initially, such perturbation induces an absorption-emission cycle
(the phenomenon of periodic oscillation of the probability for the
system being found at one of its two states) at the resonance
condition. More specifically, consider a two-state physical system
of Hamiltonian $\mathcal{H} = E_1 |1 \rangle \langle 1| + E_2 |2
\rangle \langle 2|$, with $E_2 > E_1$, in a sinusoidal potential
$\mathcal{V}(t) = \gamma e^{i \omega t} |1 \rangle \langle 2| +
\gamma e^{-i \omega t} |2 \rangle \langle 1|$. The state evolution
of the system is governed by the Schr\"{o}dinger equation
\begin{eqnarray}
\label{eqn:sch} i \frac{\partial \psi(t)}{\partial t} =
\widetilde{\mathcal{H}}(t) \psi(t)
\end{eqnarray}
with $\psi(t) = c_1(t) |1\rangle + c_2(t) |2\rangle$ and
$\widetilde{\mathcal{H}}(t) = \mathcal{H} + \mathcal{V}(t)$. For
the initial condition $c_1(0) = 0$ and $c_2(0) = 1$, the
respective probabilities of finding the system in states
$|1\rangle$ and $|2\rangle$ are of exact solutions (Rabi's
formula; see, e.g., \cite{Sakurai94})
\begin{eqnarray}
\label{eqn:c_1} |c_1(t)|^2 & = & \frac{\gamma^2}{\gamma^2 +
(\omega - \omega_{21})^2/4} \sin^2(\Omega t) \\
|c_2(t)|^2 & = & 1 - |c_1(t)|^2
\end{eqnarray}
where $\omega_{21} \equiv (E_2 - E_1)$ and $\Omega \equiv
\sqrt{\gamma^2 + \frac{(\omega - \omega_{21})^2}{4}}$. At
resonance, $\omega = \omega_{21}$, the probability of finding the
system in the ground state $|1\rangle$ oscillates with period $\pi
/ \gamma$, and reaches $1$ at time $(2k+1) \pi / (2\gamma)$, $k =
0, 1, 2, \ldots$. This phenomenon reveals the potential usefulness
of harmonic perturbation in searching the ground state of a given
Hamiltonian even with arbitrarily multiple states.

Intuitively harmonic perturbation to a general multiple-state
system may be used as a mechanism searching the ground state of
the corresponding Hamiltonian. Knowing the energy gap between the
initial state and ground state, one can apply a perturbation at
the resonance frequency to induce a probability oscillation such
that the system swings between these two states. If a measurement
is performed at the right time, the system situates definitely in
its ground state. Based on this intuition, we devise a quantum
search algorithm. An analysis shows that, given a Hamiltonian in
$N$-dimensional vector space with energy gap $E$ between its
ground state and the other $N-1$ excited states, our algorithm
finds the ground state in time $T$ with probability $\Pr = 1/(1+ c
N E^{-2} T^{-2})$ for some constant $c$. Therefore, to search the
right target with a constant high probability independent of $N$,
the time-energy product complexity matches prior known
$\Theta(\sqrt{N})$ result \cite{BBBV97} of other quantum search
algorithms \cite{Gro96,FG98,FGGS00}. Note that, due to the
time-energy duality, it is not meaningful to speak about only time
or energy complexity, regardless of the other.

\section{\label{sec:qsa}Quantum Search Algorithms Using Harmonic Perturbation}
We formulate the database search problem as follows. Given a
Hamiltonian $\mathcal{H} = \Sigma_{j=1}^{N} E_j |j \rangle \langle
j|$, we are asked to find state $|g \rangle$ such that $E_g$ is
the minimum among $E_j$'s, i.e., $|g\rangle$ is the ground state.
In the sequel, we assume any state $|j\rangle$ is a configuration
of $n$ binary digits; thus, $N = 2^n$.
Moreover, we assume that the state of the underlying physical
system is measurable such that its $n$-bit configuration (e.g.,
spin orientations of spin-$1\!/2$ particles measured along some
axis) is completely determined, and that its corresponding energy
under $\mathcal{H}$ can be observed thereafter. Unless otherwise
stated, we shall focus on the energy distribution of Grover's
search problem, and assume one out of the $N$ states is the ground
state of energy $0$ and the other $N-1$ states are excited states
of energy $E > 0$. In our algorithm, we consider $E$ is adjustable
in analyzing energy complexity.

Below we show the new application of harmonic perturbation in
quantum database search. In principle, if the perturbation
potential is designed properly, knowing the initial (excited)
state and energy gap of a given $N$-state system, one can apply
the resonance frequency to induce an oscillatory transition almost
solely between the initial state and the ground state, similar to
the two-state case. Measuring the system at the right time
achieves the highest probability of locating the ground state. How
this peak probability is related to $N$ is our main concern. To
gain an insight on the optimality limit of quantum search using
harmonic perturbation, we begin with a simple trial (in
Figure~\ref{fig:alg_nqs}) and then proceed with an optimized
procedure (in Figure~\ref{fig:alg_oqs}).

\subsection{\label{sec:qsa-trial}A trial algorithm}
\begin{figure}[t]
\vspace{-.3em}
\smallssp
\begin{program}
\>  {\bf \textit{Algorithm 1: Na\"{i}ve Ground-State Search}}\\
\> \> \INPUT:  a Hamiltonian $\mathcal{H}$\\
\> \> \OUTPUT: the ground state of $\mathcal{H}$\\
\> \> \BEGIN\\
01\ \ \= measure state, yielding say $|g_1\rangle$.\\
02     \> \IF $E_{g_1} = 0$ \ \RETURN $|g_1\rangle$\\
03     \> apply harmonic perturbation $\mathcal{V}_{[g_1,\gamma,E]}(t)$.\\
04     \> measure state at time $\pi/(2\gamma)$, yielding say $|g_2\rangle$.\\
05     \> \RETURN $|g_2\rangle$\\
\END
\end{program}
\vspace{-.9em}
\caption{The procedure of a na\"{i}ve quantum search.}
\label{fig:alg_nqs}
\end{figure}

Figure~\ref{fig:alg_nqs} sketches a (non-optimal) quantum search
procedure. The algorithm starts with a measurement in the state
basis to enforce the underlying physical system collapsing to some
state, say $|g_1\rangle$.
If the corresponding eigenenergy of the input Hamiltonian
$\mathcal{H}$ in state $|g_1\rangle$ equals $0$, the algorithm has
found the target and returns $|g_1\rangle$ immediately at Step~2.
Otherwise, harmonic perturbation is applied using the sinusoidal
potential $\mathcal{V}_{[j,\gamma,\omega]}(t)$ with
\begin{eqnarray*}
\langle p | \mathcal{V}_{[j,\gamma,\omega]}(t) | q \rangle =
\left\{
\begin{array}{ll}
\gamma e^{i \omega t}  & \mbox{if $q = j$ and $q \neq p$} \\
\gamma e^{-i \omega t} & \mbox{if $p = j$ and $p \neq q$} \\
0                      & \mbox{otherwise} \end{array} \right.
\end{eqnarray*}
for indices $p, q = 1, \ldots, N$. By replacing index $j$ with
$g_1$ and letting $\omega = E$, the perturbation
$\mathcal{V}_{[g_1,\gamma,E]}(t)$ at Step~3 induces an oscillatory
probability for the system swinging mainly between state
$|g_1\rangle$ and the unknown ground state. The algorithm
measures, at Step~4, the state of the system at time
$\pi/(2\gamma)$, when the system situates in the ground state with
the highest probability.

\begin{figure}[t]
\begin{center}
\includegraphics[height=6.8cm]{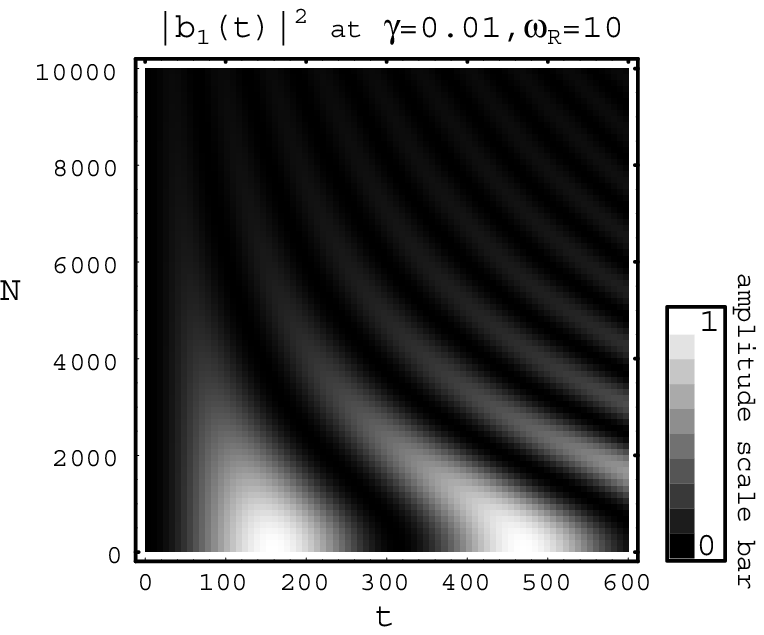}
  \caption{For $\gamma=0.01, \omega_R=10$, the
  figure plots $|b_1(t)|^2$ of Algorithm~1 under various $N$ and $t$.
  It shows that $\max_t |b_1(t)|^2$ decreases and
  $|b_1(t)|^2$ oscillates faster as $N$ gets larger.}
  \label{fig:b1t-vs-N}
\end{center}
\end{figure}

\begin{figure}[t]
\begin{center}
\includegraphics[height=6.8cm]{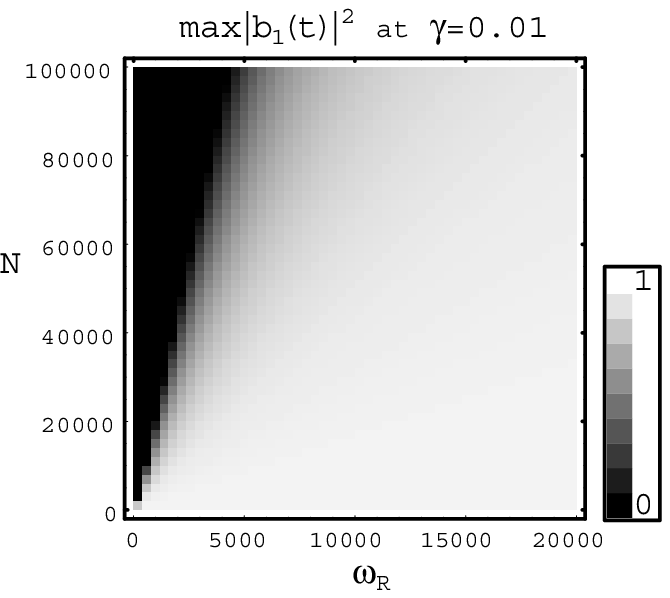}
  \caption{For $\gamma=0.01$, the figure plots
  $\max_t |b_1(t)|^2$ of Algorithm~1 under various $N$ and $\omega_R$.
  It shows that letting $\omega_R \propto N$ maintains $\max_t |b_1(t)|^2$
  at some constant value.}
  \label{fig:maxb1-vs-N-vs-w}
\end{center}
\end{figure}

We analyze the condition under which this peak probability is
independent of the effect of $N$ and is close to $1$. For $N = 2$,
the returned state $|g_2\rangle$ of the algorithm is the ground
state with certain. However, it is not the case for $N > 2$. To
see why, we solve the Schr\"{o}dinger equation of the $N$-state
system with Hamiltonian $\widetilde{\mathcal{H}}(t) = \mathcal{H}
+ \mathcal{V}_{[g_1,\gamma,E]}(t)$. To simplify the discussion,
assume without loss of generality that $|1\rangle$ is the ground
state and $|g_1\rangle = |2\rangle$ is the initial state. Let
$c_k(t)$ denote the probability amplitude of state $|k\rangle$ at
time $t$. Then the original $N$ first-order differential equations
from Equation~(\ref{eqn:sch}) can be reduced to three due to the
equivalence of $c_3(t), c_4(t), \cdots, c_N(t)$. Hence, at
resonance $\omega = E \equiv \omega_R$, the reduced equations in
terms of $b_k(t) \equiv e^{i E_k t} c_k(t)$ are
\begin{eqnarray}
i \dot{b_1}(t) & = & \gamma b_2(t) \\
i \dot{b_2}(t) & = & \gamma b_1(t) + (N-2)\gamma e^{-i \omega_R t}b_3(t)\\
i \dot{b_3}(t) & = & \gamma e^{i \omega_R t} b_2(t)
\end{eqnarray}
To solve $|b_1(t)|^2$ and thus $|c_1(t)|^2$, apply Laplace
transform $\mathcal{L}$ on these equations and solve for $B_1(s)
\equiv \mathcal{L}\{b_1(t)\}$ with initial conditions $b_2(0) = 1$
and $b_k(0) = 0$ for $k \neq 2$. We derive
\begin{eqnarray}\label{eqn:b1-t}
B_1(s) = \frac{-i\gamma}{(s^2 + \gamma^2)}\ \frac{1}{(1 +
\frac{\gamma^2}{(s^2 + \gamma^2)} \Lambda_1)}
\end{eqnarray}
where $\Lambda_1 \equiv (N-2)s/(s+i\omega_R)$. From the inverse
Laplace transform of $B_1(s)$, the exact solution of $c_1(t)$ can
be derived. For $N=2$ and thus $\Lambda_1 = 0$, $c_1(t)$ reduces
to Equation~(\ref{eqn:c_1}). For general $N > 2$, we omit listing
the sophisticated expression of $c_1(t)$ as the trial procedure is
not our final destination. Nevertheless, the numerical simulations
as shown in
Figures~\ref{fig:b1t-vs-N}~and~\ref{fig:maxb1-vs-N-vs-w} reveal
the performance of Algorithm~1 with respect to $N$. An analysis
suggests that, for a fixed constant $\gamma$ (among other
possibilities), to maintain a constant peak probability, $\Pr =
\max_t |c_1(t)|^2$, we need $\omega_R \propto N$. Thus, the
algorithm has time complexity $O(1)$ due to the fixed perturbation
amplitude $\gamma$, and has energy complexity
$O(\|H\|_2+\|\mathcal{V}\|_2) = O(\omega_R + \gamma \sqrt{N}) =
O(N)$ for $\omega_R \in O(N)$. This constant complexity in time
and linear complexity in energy can be explained in the $s$-domain
by observing that $N$ and $\omega_R$ in $\Lambda_1$ of
Equation~(\ref{eqn:b1-t}) are of a first-order relation. That is,
by maintaining $N/\omega_R = \epsilon$ for some small constant
$\epsilon$, the effect of $\Lambda_1$ is negligible and
$|c_1(t)|^2 \approx \sin^2(\gamma t)$.
Hence, the algorithm achieves the same linear resource complexity
as the classical algorithm for database search.

As a digression, we noticed that the recent work \cite{RAD06}
presented an algorithm on quantum search with resonance, where the
perturbation is very similar to
$\mathcal{V}_{[j,\gamma,\omega]}(t)$ of Algorithm~1 and thus the
algorithm is unlikely to be superior to the classical counterpart.
The misconceived quantum improvement may be due to the ignorance
of the time-energy product complexity as well as due to
over-simplified analysis \footnote{The underlying Hamiltonians for
numerical simulations in \cite{RAD06} were taken from the quantum
harmonic oscillator and the two-dimensional quantum rotor, which
have eigenenergies $E_m = E_0 (m + 1/2)$ and $E_m = E_0 m^2$,
respectively. Since these two Hamiltonians are non-degenerate, the
highest (or average) eigenenergies in consideration must be no
less than $\Omega(N)$ and $\Omega(N^2)$, respectively, which
corresponds to our notion of energy complexity. On the other hand,
the Schr\"{o}dinger equations to be solved were oversimplified and
resulted in a problematic conclusion about the complexity
improvements. In particular, since energy gaps $(E_i - E_j) \gg
1/\sqrt{N}$ for any $i \neq j$, the probability amplitude $c_k(t)$
of any state other than the initial and target states was assumed
to be 0. However, this assumption can be invalid because the
effect of $N - 2$ such small amplitudes may not be ignored.}.
Moreover, the formulation assumed that the eigenenergy of the
search target is unique and known \emph{a priori}. However, under
this assumption, the energy eigenstates could be pre-computed; the
target eigenstate can be known once the eigenenergy is specified
without even resort to quantum search. (In contrast, we assume
that the excited energy eigenstates are degenerate and there is
another observable commute with the Hamiltonian that further
determines the states.)

\subsection{A refined algorithm}
\begin{figure}[t]
\vspace{-.3em}
\smallssp
\begin{program}
\>  {\bf \textit{Algorithm 2: Optimized Ground-State Search}}\\
\> \> \INPUT:  a Hamiltonian $\mathcal{H}$\\
\> \> \OUTPUT: the ground state of $\mathcal{H}$\\
\> \> \BEGIN\\
01\ \ \= measure state, yielding say $|g_1\rangle$.\\
02     \> \IF $E_{g_1} = 0$ \ \RETURN $|g_1\rangle$\\
03     \> apply harmonic perturbation $\mathcal{V}^o_{[g_1,\gamma,E]}(t)$.\\
04     \> measure state at time $\pi/(2\gamma)$, yielding say $|g_2\rangle$.\\
05     \> \IF $E_{g_2} = 0$ \ \RETURN $|g_2\rangle$\\
06     \> apply harmonic perturbation $\mathcal{V}^e_{[g_2,\gamma,E]}(t)$.\\
07     \> measure state at time $\pi/(2\gamma)$, yielding say $|g_3\rangle$.\\
08     \> \RETURN $|g_3\rangle$\\
\END
\end{program}
\vspace{-.9em}
\caption{The procedure of an optimized quantum search.}
\label{fig:alg_oqs}
\end{figure}

Based on the insight from the $s$-domain analysis, we obtain an
optimized algorithm.
Figure~\ref{fig:alg_oqs} sketches a refined quantum search
algorithm similar to that of Figure~\ref{fig:alg_nqs}. It differs
from the procedure of Figure~\ref{fig:alg_nqs} mainly in an
additional iteration, and in the applied perturbations
$\mathcal{V}^o_{[j,\gamma,\omega]}(t)$ and
$\mathcal{V}^e_{[j,\gamma,\omega]}(t)$, where
\begin{eqnarray*}
\langle p | \mathcal{V}^o_{[j,\gamma,\omega]}(t) | q \rangle =
\left\{
\begin{array}{ll}
\gamma e^{i \omega t}  & \mbox{if ($p = j \neq q$ and $q$ even) or}\\
                       & \ \ \ \mbox{($q = j \neq p$ and $p$ odd)} \\
\gamma e^{-i \omega t} & \mbox{if ($q = j \neq p$ and $p$ even) or}\\
                       & \ \ \ \mbox{($p = j \neq q$ and $q$ odd)} \\
0                      & \mbox{otherwise} \end{array} \right.
\end{eqnarray*}
for indices $p, q = 1, \ldots, N$ and
$\mathcal{V}^e_{[j,\gamma,\omega]}(t) =
\mathcal{V}^o_{[j,\gamma,-\omega]}(t)$ (i.e.,
$\mathcal{V}^o_{[j,\gamma,\omega]}(t)$ and
$\mathcal{V}^e_{[j,\gamma,\omega]}(t)$ are conjugate to each
other). For instance, $\mathcal{V}^o_{[2,\gamma,\omega]}(t)$ in a
$4 \times 4$ matrix reads
\begin{eqnarray*}
\left(%
\begin{array}{cccccc}
  0                      & \gamma e^{ i \omega t} & 0                      & 0 \\
  \gamma e^{-i \omega t} & 0                      & \gamma e^{-i \omega t} & \gamma e^{ i \omega t} \\
  0                      & \gamma e^{ i \omega t} & 0                      & 0 \\
  0                      & \gamma e^{-i \omega t} & 0                      & 0 \\
\end{array}%
\right).
\end{eqnarray*}
Because $\mathcal{V}^o_{[j,\gamma,\omega]}(t)$ and
$\mathcal{V}^e_{[j,\gamma,\omega]}(t)$ induce noticeable
probability oscillation only when the ground state, say
$|g\rangle$, situates at an odd (i.e. $g$ odd) and even (i.e. $g$
even) position, respectively, Algorithm~2 requires one more
perturbation-and-measurement iteration than Algorithm~1. If the
ground state, not returned in Step~2 of Figure~\ref{fig:alg_oqs},
situates at an odd (respectively even) position, it will be
returned at Step~5 (respectively Step~8) almost for sure. The
algorithm can be repeated to further enhance its correctness
probability. (Notice that, in preparing $\mathcal{V}^o$ and
$\mathcal{V}^e$, we need to specify the initial state for
parameter $j$. Accordingly the initialization measurements at
Steps~1 and 4 determine the parameters $j$ of
$\mathcal{V}^o_{[j,\gamma,\omega]}(t)$ and
$\mathcal{V}^e_{[j,\gamma,\omega]}(t)$, respectively. That is why
the perturbation $\mathcal{V}^e$ of Step~6 depends on the
measurement result $|g_2\rangle$ of Step~4. Alternatively one may
discharge this dependence by fixing $j$ to some specific state.)

We analyze the condition under which the probability for
Algorithm~2 finding the search target is independent of the effect
of $N$ and is close to $1$. To simplify our discussion, again we
assume without loss of generality that $|1\rangle$ is the ground
state of $\mathcal{H}$ and the first measurement in Step~1 of
Figure~\ref{fig:alg_oqs} yields $|g_1\rangle = |2\rangle$. To
compute the probability that the ground state $|1\rangle$ is
correctly returned in Step~5, we solve the Schr\"{o}dinger
equation of an $N$-state system with Hamiltonian
$\widetilde{\mathcal{H}}(t) = \mathcal{H} +
\mathcal{V}^o_{[2,\gamma,E]}(t)$. Let $c_k(t)$ be the probability
amplitude of state $|k\rangle$ at time $t$. Then the original $N$
first-order differential equations can be reduced to four due to
the equivalence of odd coefficients $c_3(t),\ldots,c_{N-1}(t)$ and
the equivalence of even coefficients $c_4(t),\ldots,c_N(t)$.
Hence, at resonance $\omega = E \equiv \omega_R$, the reduced
equations in terms of $b_k(t) \equiv e^{i E_k t} c_k(t)$ are
\begin{eqnarray}
\label{eqn:ob1} i \dot{b_1}(t) & = & \gamma b_2(t) \\
\label{eqn:ob2} i \dot{b_2}(t) & = & \gamma b_1(t) + \nonumber \\  & & 
\frac{(N-2)}{2}\gamma (e^{-i \omega_R t} b_3(t)
 + e^{i \omega_R t} b_4(t))\\
\label{eqn:ob3} i \dot{b_3}(t) & = & \gamma e^{i \omega_R t} b_2(t) \\
\label{eqn:ob4} i \dot{b_4}(t) & = & \gamma e^{-i \omega_R t}
b_2(t)
\end{eqnarray}
To solve $|b_1(t)|^2$ and thus $|c_1(t)|^2$, we apply Laplace
transform $\mathcal{L}$ on these equations and solve for $B_1(s)
\equiv \mathcal{L}\{b_1(t)\}$ with initial conditions $b_2(0) = 1$
and $b_k(0) = 0$, $k \neq 2$. We derive
\begin{eqnarray}\label{eqn:b1-o}
B_1(s) = \frac{-i\gamma}{(s^2 + \gamma^2)}\ \frac{1}{(1 +
\frac{\gamma^2}{(s^2 + \gamma^2)} \Lambda_2)}
\end{eqnarray}
where $\Lambda_2 \equiv (N-2)s^2/(s^2+ \omega_{R}\!^2)$. Taking
inverse Laplace transform $\mathcal{L}^{-1}\{ B_1(s)\}$ and
assuming $N \gg 1$, we get
\begin{eqnarray}\label{eqn:main}
b_1(t) \approx \frac{-i\omega_R}{\sqrt{N \gamma^2 + \omega_R\!^2}}
\sin\left(\frac{\gamma\omega_R}{\sqrt{N\gamma^2 + \omega_R\!^2}}
t\right)
\end{eqnarray}
By Equation~(\ref{eqn:main}), the peak probability $\Pr$ of
finding the search target equals $\max_t |c_1(t)|^2 = 1/(1+ N
\gamma^2 \omega_R\!^{-2})$; the period of the probability
oscillation of $|c_1(t)|^2$ is $\tau = \pi \sqrt{1 + N\gamma^2
\omega_R\!^{-2}} / \gamma$. Since on average the search target can
be found by running the algorithm $1/\Pr$ times each of which
takes time $\tau$, the total time complexity is of $\tau/\Pr \in
O((1 + N\gamma^2 \omega_R\!^{-2})^{\frac{3}{2}} / \gamma)$. On the
other hand, the energy complexity is of
$O(\|H\|_2+\|\mathcal{V}\|_2) = O(\omega_R + \gamma \sqrt{N})$. It
can be verified that maintaining a constant $\Pr$ achieves the
tightest upper bound of resource complexity.
As a result, when $\Pr$ is maintained as a constant (as we shall
assume in the sequel), $\gamma$ is the only parameter affecting
time complexity $T$. Letting $E$ denote the energy complexity, we
can write $\Pr = 1/(1+ c N E^{-2} T^{-2})$ for some constant $c$.
In essence, the time-energy product complexity is of
$O(\sqrt{N})$.

The foregoing analysis assumes that the ground state is in an odd
position. Suppose that the ground state is rather in some even
position. Then $\mathcal{V}^o_{[g_1,\gamma,E]}(t)$ of Step~3 does
not induce a state oscillation (and thus $|g_2\rangle$ is the same
as $|g_1\rangle$ with high probability) while
$\mathcal{V}^e_{[g_2,\gamma,E]}(t)$ of Step~6 does. In this case,
a similar analysis holds to derive the probability that the
algorithm returns a correct answer at Step~8.

Since Algorithm~2 achieves the $O(\sqrt{N})$ time-energy product
complexity, it is optimal according to the analysis of
\cite{FG98}. Below we provide yet another proof using $s$-domain
analysis. It may provide an intuition why Algorithm~2 outperforms
Algorithm~1. Also we hope that the technique of $s$-domain
analysis may become a powerful tool for the complexity analysis of
other quantum algorithms. As least in our cases, $s$-domain
analysis is more intuitive than time-domain analysis.

\section{\label{sec:ol}The Optimality Limit}
Comparing Algorithms~1~and~2, one might think that the performed
modification might seem suspiciously irrelevant as it only affects
states $|3\rangle, \ldots, |N\rangle$ in our example. However,
this is not true because it in fact affects $|2\rangle$ by
Equation~(\ref{eqn:ob2}) and thus $|1\rangle$ by
Equation~(\ref{eqn:ob1}). (Note that Algorithm~2 on perturbation
$\mathcal{V}^o$, and also $\mathcal{V}^e$, applies both $e^{i
\omega t}$ and $e^{-i \omega t}$ simultaneously, rather than
sequentially. It differs from applying $e^{i \omega t}$ on some
states for one search and then applying $e^{-i \omega t}$ on other
states for another search. Otherwise, Algorithm~2 would be
equivalent to performing Algorithm~1 for several times and has no
improvements.) Although the effect may seem obscure from the
time-domain Schr\"{o}dinger equations, it is apparent when solving
them in the $s$-domain. The two different phases in
$\mathcal{V}^o$ and $\mathcal{V}^e$ of Algorithm~2 may interfere
and simplify $B_1(s)$.

To understand the improvement of Algorithm~2 over Algorithm~1,
compare Equations~(\ref{eqn:b1-t}) and (\ref{eqn:b1-o}). We see
that the undesirable effect of large $N$ in $\Lambda_2$ is
nullified by $\omega_R\!^2$ while that in $\Lambda_1$ is nullified
by $\omega_R$. As a consequence, for constant $\gamma$, Algorithms
1 and 2 are of energy complexities $O(N)$ and $O(\sqrt{N})$,
respectively. Ideally, if we can construct a $\Lambda$ such that
the effect of $N$ is cancelled out by $\omega_R\!^d$ for a larger
exponent $d$, then $\Pr$ can be maintained as a constant with a
lower resource complexity. Unfortunately, we will now show that
$d$ is at most $2$, that is, no improvement is possible by
introducing more different phases and/or amplitudes to the
non-zero entries of $\mathcal{V}^{o}_{[j,\gamma,\omega]}(t)$ and
$\mathcal{V}^{e}_{[j,\gamma,\omega]}(t)$. (Note that replacing the
zero entries of $\mathcal{V}^{o}_{[j,\gamma,\omega]}(t)$ and
$\mathcal{V}^{e}_{[j,\gamma,\omega]}(t)$ with non-zero elements
introduces undesirable amplitude leakage to states other than the
initial and target ground states. Hence we only need to consider
modifying the non-zero entries.)

Consider a general perturbation potential (a Hermitian matrix)
with non-vanishing entries only in the row and column indexed by
some initial state $|j\rangle$ similar to
$\mathcal{V}_{[j,\gamma,\omega]}(t)$. These entries can have
arbitrary amplitudes and frequencies. (However, one of the
frequencies must equal $\omega_R$ such that resonance is
possible.) Assuming without loss of generality $|1\rangle$ and
$|2\rangle$ to be the ground and initial state, respectively, we
solve the corresponding Schr\"{o}dinger equations and can write
\begin{eqnarray*}
B_1(s) = \frac{-i\gamma}{(s^2 + \gamma^2)}\ \frac{1}{(1 +
\frac{\gamma^2}{(s^2 + \gamma^2)} \Lambda)}
\end{eqnarray*}
with $\gamma$ being the amplitudes for entries $|1\rangle\langle
2|$ and $|2\rangle\langle 1|$, and
\begin{eqnarray}\label{eqn:lambda}
\Lambda & \equiv & s\left(\frac{\alpha_1}{s+i\omega_1} +
\frac{\alpha_2}{s+i\omega_2} + \cdots +
\frac{\alpha_m}{s+i\omega_m}\right)
\end{eqnarray}
where index $m$ is polynomial in $n$ (thus $m \ll N = 2^n$),
$\alpha_j$'s are positive real numbers, $\omega_j$'s are of the
form $a_j \omega_R\!^{x_j}$ for real constants $a_j$ and $x_j$,
and $\omega_j \neq \omega_k$ for $j \neq k$. (Note that $\alpha_j$
is obtained from the product of some complex number and its
complex conjugate, and thus is positive.)
\begin{theorem} \label{thm:opt}
Let $d$ equal the largest exponent in terms of $\omega_R$ in the
denominator of Equation~(\ref{eqn:lambda}) minus that in the
nominator. Then, $d \leq 2$ for all possible assignments to the
constant parameters $\alpha_j$, $a_j$, and $x_j$ of $\Lambda$.
\end{theorem}
Since $d = 2$ is achieved, the quantum search algorithm of
Figure~\ref{fig:alg_oqs} is optimal.

\section{\label{sec:disc}Discussions}
We compare various quantum search algorithms, namely, those of
\cite{Gro96,FG98,FGGS00} and ours, in terms of the generalized
problem that the search target appears more than once in the
database (i.e., there are multiple ground states in our setup).
For \cite{Gro96}, the number of appearances needs to be known
\textit{a priori} for accurate search (unless more complicated
quantum counting \cite{BBHT98} is incorporated); for \cite{FG98},
the method may also seem no easy extension. In contrast, adiabatic
computation \cite{FGGS00} supports multiple appearances of search
target. However, it requires careful analysis to decide valid
evolution speed; whether there is quantum speedup may depend on
the instance to be solved. On the other hand, our approach, when
applied to the generalized problem, may not always yield quantum
speedup. An analysis shows that a large number of ground
odd-states (respectively even-states) for perturbation
$\mathcal{V}^o$ (respectively $\mathcal{V}^e$) makes the
multiplicities of $b_3(t)$ and $b_4(t)$ in
Equation~(\ref{eqn:ob2}) differ to some extent. Thus the advantage
of Algorithm~2 with respect to Algorithm~1 disappears. It would be
interesting to know if our approach can be extended to handle this
generalized problem with definite quantum speedup. Also whether
our method has unseen advantages over other methods remains
further investigation.

Even though our perturbation Hamiltonian looks more complicated
than that of \cite{FG98}, it does not directly imply our proposal
is harder to implement. The implementation issue is beyond the
scope of our focus and poses a challenge to the experimentalists.
Nevertheless, it at least suggests a new way of performing quantum
search.

\section{\label{sec:cncl}Conclusions}
Motivated by developing a quantum search algorithm, we formulated
a multi-state harmonic perturbation problem and gave an exact
analysis. This paper answered affirmatively and constructively the
open question whether fast time-varying Hamiltonians can be
exploited in quantum search. The presented algorithm is optimal
and achieves a quadratic speed-up over classical algorithms
similar to prior methods \cite{Gro96,FG98,FGGS00}. Under this new
computation model, we hope it may suggest different approaches to
the realization of quantum computers.


\bibliographystyle{unsrt}
\bibliography{qshp-qip}

\begin{thebibliography}{10}

\bibitem{Gro96}
L.~Grover.
\newblock A fast quantum mechanical algorithm for database search.
\newblock In {\em Proc. the 28th ACM Symposium on the Theory of Computing},
  pages 212--219, 1996.

\bibitem{FG98}
E.~Farhi and S.~Gutmann.
\newblock Analog analogue of a digital quantum computation.
\newblock {\em Physical Review A}, 57(4):2403--2406, 1998.

\bibitem{FGGS00}
E.~Farhi, J.~Goldstone, S.~Gutmann, and M.~Sipser.
\newblock Quantum computation by adiabatic evolution.
\newblock quant-ph/0001106, 2000.

\bibitem{DMV01}
W.~van Dam, M.~Mosca, and U.~Vazirani.
\newblock How powerful is adiabatic quantum computation?
\newblock In {\em Proc. the 42th IEEE Symposium on Foundations of Computer
  Science}, pages 279--287, 2001.

\bibitem{RC02}
J.~Roland and N.~Cerf.
\newblock Quantum search by local adiabatic evolution.
\newblock {\em Physical Review A.}, 65(4):042308(6), 2002.

\bibitem{ADK04}
D.~Aharonov, W.~van Dam, J.~Kempe, Z.~Landau, S.~Lloyd, and O.~Regev.
\newblock Adiabatic quantum computation is equivalent to standard quantum
  computation.
\newblock In {\em Proc. the 45th IEEE Symposium on Foundations of Computer
  Science}, 2004.

\bibitem{VSS00}
L.~Vandersypen, M.~Steffen, M.~Sherwood, C.~Yannoni, G.~Breyta, and I.~Chuang.
\newblock Implementation of a three-quantum-bit search algorithm.
\newblock {\em Applied Physics Letters}, 76(5):646--648, 2000.

\bibitem{PZFFLG02}
X.~Peng, X.~Zhu, X.~Fang, M.~Feng, M.~Liu, and K.~Gao.
\newblock Experimental implementation of \textsc{H}ogg's algorithm on a
  three-quantum-bit \textsc{NMR} quantum computer.
\newblock {\em Physical Review A}, 65:042315, 2002.

\bibitem{SvD03}
M.~Steffen, W.~van Dam, T.~Hogg, G.~Breyta, and I.~Chuang.
\newblock Experimental implementation of an adiabatic quantum optimization
  algorithm.
\newblock {\em Physical Review Letters}, 90(6):067903(4), 2003.

\bibitem{Pre98}
J.~Preskill.
\newblock Quantum computing: pro and con.
\newblock {\em Proc. R. Soc. Lond. A}, 454:469--486, 1998.

\bibitem{Sakurai94}
J.~J. Sakurai.
\newblock {\em Modern Quantum Mechanics}.
\newblock Addison Wesley, 1994.

\bibitem{BBBV97}
C.~Bennett, E.~Bernstein, G.~Brassard, and U.~Vazirani.
\newblock Strengths and weaknesses of quantum computation.
\newblock {\em Siam Journal of Computing}, 26:1277--1339, 1997.

\bibitem{RAD06}
A.~Romanelli, A.~Auyuanet, and R.~Donangelo.
\newblock Quantum search with resonances.
\newblock {\em Physica A.}, 360:274--284, 2006.

\bibitem{BBHT98}
M.~Boyer, G.~Brassard, P.~Hoeyer, and A.~Tapp.
\newblock Tight bounds on quantum searching.
\newblock {\em Fortschritte der Physik}, 46:493--506, 1998.

\end{thebibliography}

\noindent {\bf Appendix}\\
Theorem~\ref{thm:opt}\begin{proof}\\
Expanding $\Lambda$ of Equation~(\ref{eqn:lambda}), we have
\begin{eqnarray} \nonumber
\lefteqn{\Lambda = \frac{1}{(s+i\omega_1) \cdots (s+i\omega_m)} \{
s^m + \cdots +} \\
&& ( \sum_{j} \alpha_j \sum_{k \neq j} ( \prod_{l \neq j, l\neq k}
\omega_l ) ) s^2 + ( \sum_{j} \alpha_j \prod_{k \neq j} \omega_k )
s \}. \label{eqn:lambda-expand}
\end{eqnarray}
We first show that, if $d > 0$, the largest exponent in terms of
$\omega_R$ in the denominator of
Equation~(\ref{eqn:lambda-expand}) must result from the product of
all $\omega_j$'s. Let $\sigma = x_1 + \cdots + x_m$; then the
product term $\omega_1 \omega_2 \cdots \omega_m$ equals $a
\omega_{R}\!^\sigma$ for some constant $a \neq 0$. For
contradiction, suppose that the product term $\omega_1 \omega_2
\cdots \omega_m$ is not the only highest order term in $\omega_R$.
Then there must exist at least one $\omega_j = a_j
{\omega_R}^{x_j}$ with $x_j \leq 0$. Let $a^\prime
{\omega_R}^{\sigma^\prime}$ for $a^\prime \neq 0$ be the product
term of all $\omega_j$'s with $x_j > 0$. Because $a^\prime
{\omega_R}^{\sigma^\prime}$ must appear somewhere in both the
denominator and nominator of Equation~(\ref{eqn:lambda-expand}),
$d$ must equal $0$. Hence, to have $d
> 0$, all $\omega_j$'s must have $x_j > 0$. That is, the product
term $\omega_1 \omega_2 \cdots \omega_m$ contributes to the
largest exponent of $\omega_R$ in the denominator.

To have $d > 2$, the exponent of $\omega_R$ in the nominator of
Equation~(\ref{eqn:lambda-expand}) cannot be too large. Observe
that, for the resonance condition to hold in harmonic
perturbation, there must exist some $x_k = 1$. Moreover, there are
more than one such $x_k$'s. Otherwise, the product term $\omega_1
\cdots \omega_{k-1} \omega_{k+1} \cdots \omega_m$ cannot be
cancelled out. Its existence in the nominator of
Equation~\ref{eqn:lambda-expand} results in $d \leq 1$, which
violates the desired condition. Because of these constraints and
$x_j > 0$ for $j = 1,\ldots,m$, the coefficients of $s$ and $s^2$
in the nominator of Equation~(\ref{eqn:lambda-expand}) must equal
zero. That is,
\begin{eqnarray}
\label{eqn:m-1} \sum_{j} \alpha_j \prod_{k \neq j} \omega_k = 0, \mbox{\ and} \\
\label{eqn:m-2} \sum_{j} \alpha_j  \sum_{k \neq j} \left( \prod_{l
\neq j, l\neq k} \omega_l \right) = 0
\end{eqnarray}
must be satisfied. Multiplying (\ref{eqn:m-1}) by
$(\frac{1}{\omega_1} + \cdots + \frac{1}{\omega_m})$ yields
\begin{eqnarray}\label{eqn:mm}
\sum_j \alpha_j \frac{\prod_{k \neq j} \omega_k}{\omega_j} +
\sum_{j} \alpha_j \sum_{k \neq j} \left( \prod_{l \neq j, l\neq k}
\omega_l\right) = 0.
\end{eqnarray}
By recognizing that the second term of Equation~(\ref{eqn:mm})
equals $0$ by Equation~(\ref{eqn:m-2}), it is immediate that
\begin{eqnarray}\label{eqn:mms}
\sum_j \alpha_j \frac{\prod_{k \neq j} \omega_k}{\omega_j} = 0.
\end{eqnarray}
Again, multiplying Equation~(\ref{eqn:mms}) by the product term
$\omega_1 \omega_2 \cdots \omega_m$ yields
\begin{eqnarray} \label{eqn:mmf}
\sum_{j} \alpha_j \prod_{k \neq j} \omega_{k}\!^2 = 0.
\end{eqnarray}
Since $\alpha_j$'s are positive real numbers, the left-hand side
of Equation~(\ref{eqn:mmf}) must be greater than zero unless all
$\omega_j$'s equal zero, which violates the condition that
$\omega_j \neq \omega_k$ for $j \neq k$. Thus,
Equation~(\ref{eqn:mmf}) does not hold, and $d \not > 2$.
Moreover, since $d = 2$ is achieved by Algorithm 2, the theorem
follows.
\end{proof}

\end{document}